\documentclass[a4paper,11pt]{article}
\usepackage{amsfonts}

\usepackage{graphicx}
\graphicspath{{Figures/}}
\usepackage{amsmath}
\usepackage{amssymb}
\usepackage{latexsym}
\usepackage[colorlinks]{hyperref}
\usepackage{color}
\usepackage{float}
\usepackage{cite}
\usepackage{makeidx}
\usepackage{colortbl}
\usepackage{csquotes}
\usepackage[squaren]{SIunits}

\usepackage[textfont={footnotesize,sf},labelfont={color=blue,bf,sf},labelsep=endash]{caption}
\usepackage[position=top,labelfont={color=blue,bf,sf}]{subfig}
\usepackage{bm}

\usepackage[title]{appendix}
\newcommand{\bra}{\begin{array}}
	\newcommand{\era}{\end{array}}
\newcommand{\beq}{\begin{equation}}
	\newcommand{\eeq}{\end{equation}}
\newcommand{\bqr}{\begin{eqnarray}}
	\newcommand{\eqr}{\end{eqnarray}}

\begin{document}
	\begin{titlepage}
		\setcounter{page}{1}
		\renewcommand{\thefootnote}{\fnsymbol{footnote}}

		
		\begin{center}
			
			{\Large \bf {Tunneling Effect in   
					Gapped Phosphorene through  Double Barriers}}
			
			\vspace{5mm} {\bf Jilali Seffadi}$^{a}$,
			{\bf Ilham Redouani}$^{a}$, {\bf Youness Zahidi \footnote{\sf
					y.zahidi@usms.ma}}$^{b}$ and {\bf Ahmed Jellal\footnote{\sf
					a.jellal@ucd.ac.ma}}$^{a,c}$
			
			\vspace{5mm}
			
			{$^{a}$\em Laboratory of Theoretical Physics,  
				Faculty of Sciences, Choua\"ib Doukkali University},\\
			{\em PO Box 20, 24000 El Jadida, Morocco}

			{$^b$\em MRI Labortory, Polydisciplinary Faculty, Sultan Moulay Selimane
				University,\\
				PO Box 145, 25000 Khouribga, Morocco}
			
			{$^{c}$\em Canadian Quantum  Research Center,
				204-3002 32 Ave Vernon, \\ BC V1T 2L7,  Canada}

			\vspace{30mm}

			\begin{abstract}
				
We study the transport properties of charge carriers in phosphorene with a mass term through double
barriers. The solutions of the energy spectrum are obtained and the dependence of the eigenvalues on
the barrier potentials and wave vectors in the $x$-direction is numerically computed. Using the boundary
conditions together with the matrix transfer method, we determine transmission and the conductance of
our system. These two quantities are analyzed by studying their main characteristics as a function
of the physical parameters along the armchair direction. Our results show the highly anisotropic
character of phosphorene and the no signature of Klein tunneling at normal incidence contrary to
graphene. Moreover, it is found that the transmission and conductance display oscillatory behaviors
in terms of the barrier width under suitable conditions.

				\vspace{3cm}
				
					

			\end{abstract}
		\end{center}
	\end{titlepage}

	\section{Introduction}
	
	 Two dimensional (2D) crystals consisting of single or a few atomic layers has been the focus of instance research currently, due to their fundamental properties and possible applications. Since the first experimental fabrication of  graphene\cite{ref01}, a 2D single sheet of carbon honeycomb, 
the researchers were focused on it because of	 
	 its unique electronic \cite{ref1,ref2,ref02,ref002}, optical \cite{ref3} and mechanical properties \cite{ref03,ref003}. From then, there is  a growing interest in the realization of other (2D) materials. The investigation of analogs of graphene has resulted in the discovery of various single-layer crystals of different elements. Among them,  silicon (silicene) \cite{ref4}, germanium (germanene)\cite{ref5} and a class of materials known as transition-metal dichalcogenides\cite{ref6}. 
	 In recent years, it is showed that 
	 such materials serve as test bed for Dirac physics, which 
	 arises from the fact that their low-energy quasiparticles obey an effective Dirac-like equation. These quasiparticles lead to a host of unconventional transport properties and distinguishes them from conventional materials obeying the Schr\"odinger equation.
	
The black phosphorus  is 
	among the most promising  2D materials,  which is an allotrope of phosphorus\cite{ref8,ref9,ref10,ref11} and well-known as phosphorene. It has triggered tremendous attention since its first discovery in 2014 \cite{ref10,ref11}. Similarly to graphene, phosphorene can be mechanically exfoliated to obtain samples with a single or few   layers. It features an orthorhombic crystal structure, whose individual layers feature a puckered, honeycomb lattice structure and each atom is covalently bonded to three of its neighbor. Phosphorene is a semiconductor and has a high electronic mobility in 
the range of $1000$  cm$^2$ V$^{-1}$ s$^{-1}$ \cite{ref10}, which 	
	 makes it as a possible candidate for device applications\cite{ref011,ref12}.
	 Its  electronic structure  \cite{ref012} shows, in contrast to graphene, an intrinsic band-gap and strong anisotropy, which causes the electrons to behave in one direction like massive Dirac fermions and in the orthogonal direction like non-relativistic Schr\"odinger electrons \cite{ref14}. Phosphorene shows a robust direct band gap for all thickness varying from $1.8$  eV for single layer  to $0.4$  eV for bulk samples \cite{ref10}, which depends on the number of layers.
	
	The quantum transport in phosphorene was analyzed under different circumstances. For instance, it was 
	highlighted the anisotropic
	transport signatures of phosphorene across a NBN junction
	with different orientations \cite{ref17}. Additionally, it was shown
	that for a barrier along the $x$-direction, the dominant 
	contribution to the transport comes from the quasiparticles,
	 which
	 impinge on the barrier at near-normal incidence provided the
	 applied voltage is close to the bottom of the conduction band,
	 leading to collimated transport of electrons. Recently, 
	it was
	demonstrated that phosphorene PNP junctions constitute perfect
	electron waveguides, without any leakage current through the side-
	walls \cite{Betancur-Ocampo2020} and at the interfaces of the junction, the omni-directional total reflection takes place,
	which  is not due to an energetically forbidden region but due to pseudo-spin blocking.

	Motivated by the results developed in \cite{ref17, Betancur-Ocampo2020},
we consider a monolayer phosphorene and  study the scattering of charge carriers through double barrier structure with an energy gap in the central region. We solve the eigenvalue equation of the continuum model and then determine 
the solutions of energy spectrum in different regions. By requiring the continuity of the wave functions at interfaces and using the transfer matrix method, we determine the transmission and the conductance as a function of the physical parameters. 
Numerically we show that
 the transmission displays sharp pics inside
the transmission gap around some energy values, which are resulted from
  the quasibound states formed in the double barrier structure. Additionally, our analysis tells us there is no 
 signature of the Klein tunneling at normal incidence.
 Moreover, we find that 
  the conductance displays oscillatory
 behavior as a function of the barrier width.

	The present paper is organized as follows. In section \text{\ref{Sec2}},
	we describe our system by setting the appropriate Hamiltonian, which will be  used to obtain the eigenvalues and eigenspinors.
	By applying  the boundary conditions at interfaces,
	we determine the transmission probability 
	and the associated conductance. 
	In section \text{\ref{Sec3}}, we numerically analyze our results under suitable conditions of the physical parameters. Finally, we conclude our work.

	\section{Hamiltonian formalism}\label{Sec2}

\begin{figure}[h!]
		\centering
		\centering{\includegraphics[width=3.5in]{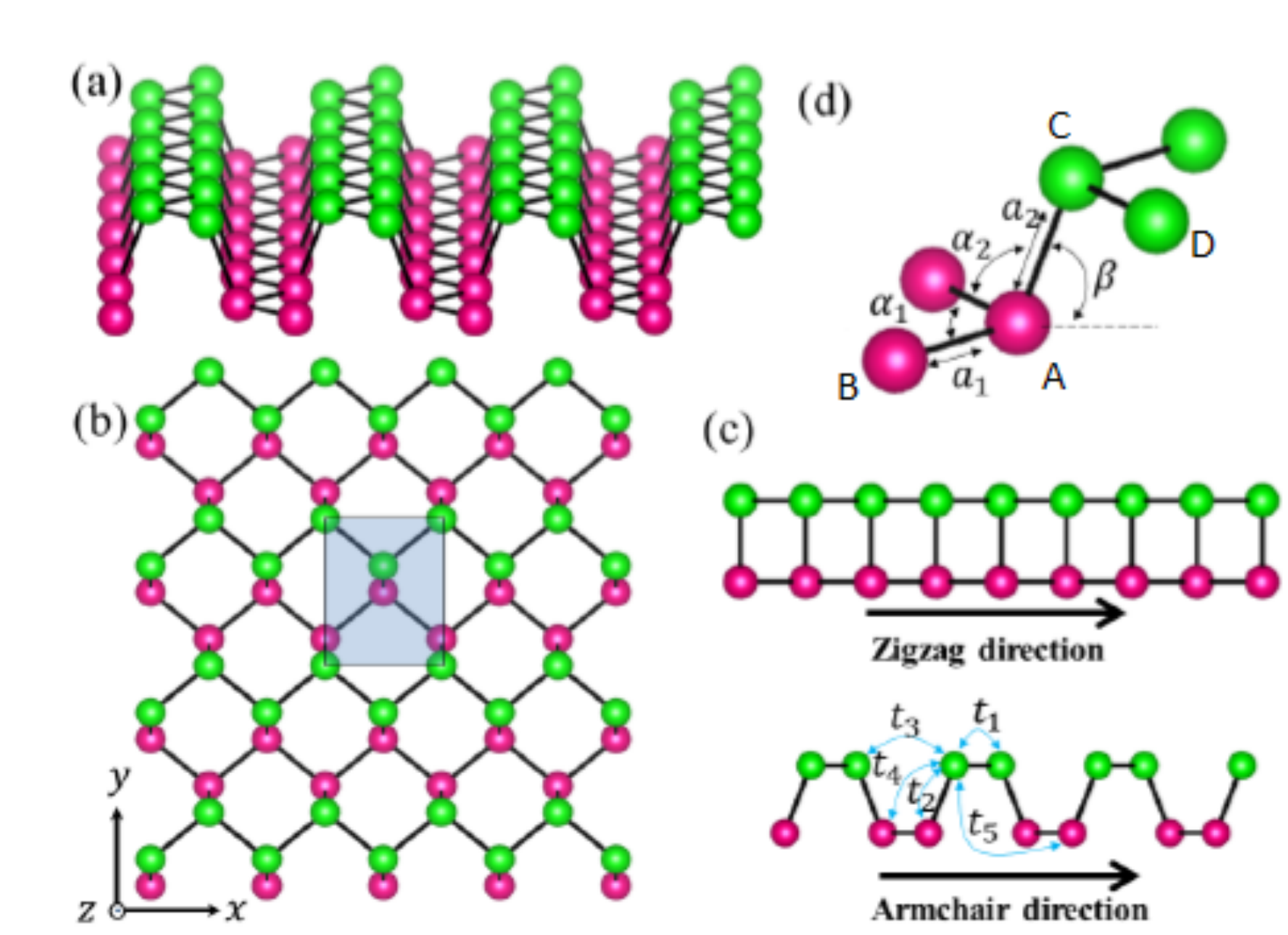}}
		\caption{\sf{ (a): Bird eye’s view of the lattice structure of monolayer phosphorene, where the phosphorus atoms at different sublayers are represented by different colors in each sublayer. (b): 2D lattice of monolayer projected onto the xy-plane. Shaded rectangle indicates the unit cell. (c): The ”zigzag” and ”armchair” edges with inter-atomic coupling. (d): Lattice structure of monolayer phosphorene systems with the different parameters.
		}}\label{f05} 
	\end{figure}

 Recall that phosphorene is a 2D hexagonal lattice  that is buckled (or "puckered") due to the  hybridization $sp^3$. The phosphorus atoms at different sublayers are represented  
 in Figure \ref{f05} with sublattices $(A, B)$ at bottom sublayer
 (pink symbol) and 
  ($C,D$) at top one  (green symbol). This geometry results in two types of bands
  as shown  as in Figure \ref{f05}(a)  where  atoms $P$,  connected by bonds parallel to the  plane, form upper and lower sublayers, while the bonds connecting  atoms $P$ between these two sublayers are oriented by an angle $71.7^{\circ}$ out of  plane. Figures \ref{f05}(b,c) present the top and side-view, along the armchair direction, of the lattice structure of monolayer phosphorene.  The shaded rectangle indicates the unit cell and the side view highlights the inter-atomic coupling. The hopping parameter $t_1$ 
  is for the connection along a zigzag direction in the upper or lower sublayer and $t_2$ is for that 
  between a pair of zigzag chains in the upper and lower sublayers. The parameters  $t_3$ and $t_4$ are between the nearest-neighbor and next nearest neighbor sites of a pair of zigzag chains in the upper or lower sublayer. The parameter $t_5$ is  between two atoms on upper and lower zigzag chains that are farthest from each other. Figure \ref{f05}(d) shows the lattice structure of phosphorene, emphasizing the bond lengths and bond angles between the phosphorus atomic sites, with $a_1 = 2.22 \ \AA$ is the distance between nearest neighbour sites in sublattices ($A, B$) or ($C, D$) and $a_2 = 2.24 \ \AA$ is that 
  between nearest neighbor sites in sublattices ($A, C$) or ($B,D$). The bond angles are $\alpha_1 = 96.5^{\circ}$, $\alpha_2 = 101.9^{\circ}$ and $\beta = 72^{\circ}$.

\begin{figure}[h!]
		\centering
		\includegraphics[width=4.5in]{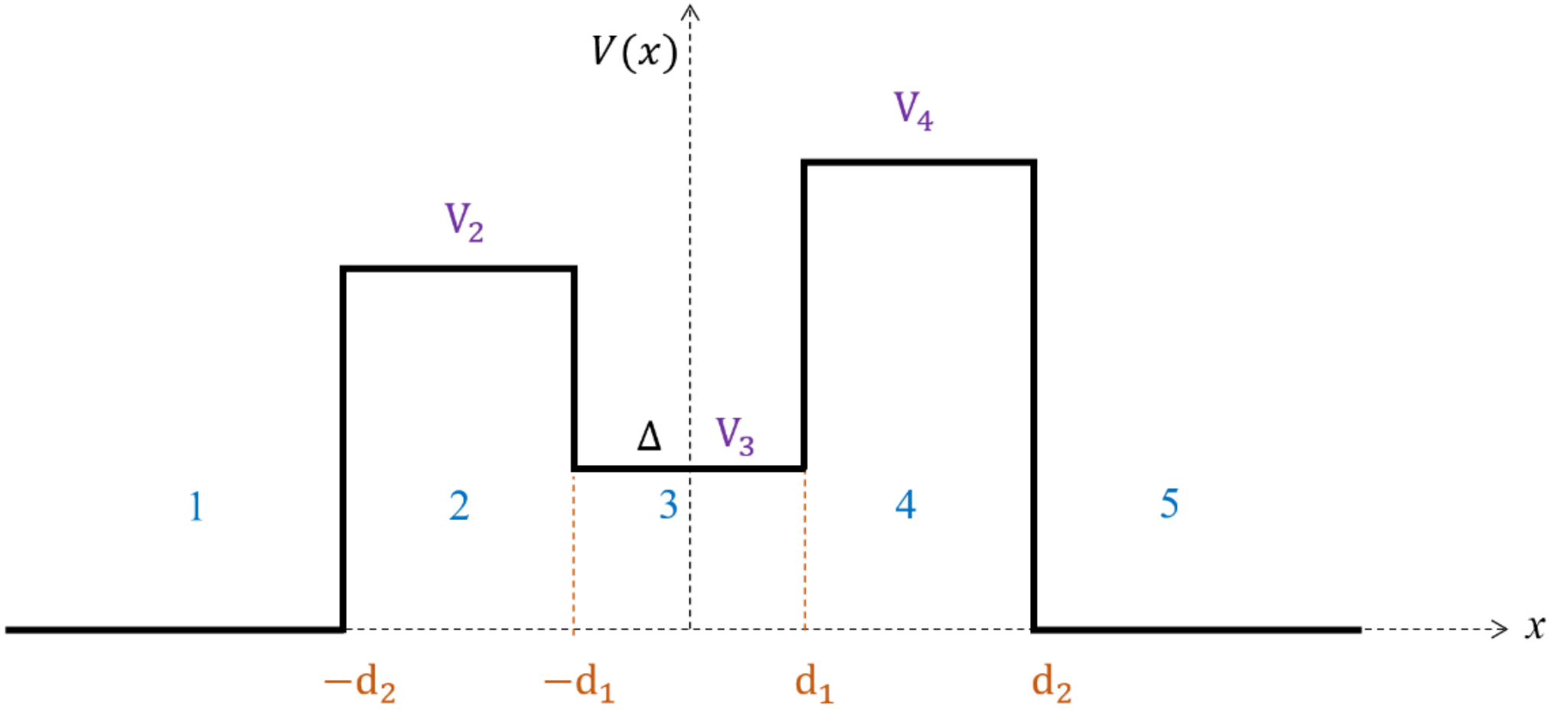}
		\caption{\sf{ Schematic profile of the barrier potentials $V(x)$ composed of five regions with three amplitudes $(V_2,V_3,V_4)$ and a gap placed in region $3$.}}\label{pprofil}
	\end{figure}

	It has been recently shown within the five-hopping parameter approach \cite{ref15,ref16} that the continuum approximation is very suited for describing the physics of large phosphorene, yielding very accurate results within its limit of validation. A unitary transformation can be performed to rewrite the monolayer Hamiltonian in a simpler block form. In our case, we consider the block Hamiltonian \cite{ref15} 
	\begin{equation}
		H^\pm(k)=\left(%
		\begin{array}{cc}
			t_{AA}(k) \pm t_{AD}(k) & 	t_{AB}(k) \pm t_{AC}(k) \\
			t_{AB}^*(k) \pm t_{AC}^*(k) & t_{AA}(k) \pm t_{AD}(k) \\
		\end{array}%
		\right)
	\end{equation}
	where under expansion around the long-wavelength $k = 0$ ($\Gamma$ point) up to second order, the structure factors become
	\begin{equation}
		t_{AA}(k)=\delta_{AA}+\eta_{AA} k_{y}^{2}+\gamma_{AA} k_{x}^{2}  \qquad\qquad\quad
	\end{equation}
	\begin{equation}
		t_{AB}(k)=\delta_{AB}+\eta_{AB} k_{y}^{2}+\gamma_{AB} k_{x}^{2}+i\chi_{AB} k_x
	\end{equation}
	\begin{equation}
		t_{AC}(k)=\delta_{AC}+\eta_{AC} k_{y}^{2}+\gamma_{AC} k_{x}^{2}+i\chi_{AC} k_x 
	\end{equation}
	\begin{equation}	
		t_{AD}(k)=\delta_{AD}+\eta_{AD} k_{y}^{2}+\gamma_{AD}  \qquad\qquad\quad
	\end{equation}
	giving rise to the following Hamiltonian 
	\begin{equation}
		H^+(k)=\left(%
		\begin{array}{cc}
			u_0+\eta_{x} k_{x}^{2}+\eta_{y} k_{y}^{2} & 		\delta+\gamma_{x} k_{x}^{2}+\gamma_{y} k_{y}^{2}+i\chi k_x \\
			\delta+\gamma_{x} k_{x}^{2}+\gamma_{y} k_{y}^{2}-i\chi k_x &	u_0+\eta_{x} k_{x}^{2}+\eta_{y} k_{y}^{2} \\
		\end{array}%
		\right)
	\end{equation}
	with the setting 
	\begin{align}
	&u_{0}=\delta_{AA}+ \delta_{AD},\qquad  \delta=\delta_{AB}+ \delta_{AC}\\
	 &\eta_{x}=\eta_{AA}+ \eta_{AD}, \qquad \eta_{y}=\gamma_{AA}+ \gamma_{AD}\\ &\gamma_{x}=\eta_{AB}+\eta_{AC}, \qquad \gamma_{y}=\gamma_{AB}+\gamma_{AC}\\
	 & 
	\chi=\chi_{AB}+ \chi_{AC}
	\end{align}
	  $\vec k=(k_x,k_y)$ is the wave vector and
	the coefficient values of the
	expanded structure factors for both five-hopping models are summarized in Table \ref{bc}.
	\begin{table}[htbp]
		\center
		\begin{tabular}{ccccc}
			\hline
			& 5-hopping    &  &5-hopping& Units \\
			\hline
			$\delta_{AA} $& 0.00& $\delta_{AB} $ &-2.85& eV \\
			$\delta_{AC} $&  3.61   &$\delta_{AD} $&-0.42 & eV \\
			$\eta_{AA}   $&  0.00   &$\eta_{AB}   $&3.91  & eV ${\AA}^{2}$ \\
			$\eta_{AC}   $&-0.53 & $\eta_{AD}   $&0.58 & eV ${\AA}^{2}$ \\
			$\gamma_{AA}$ & 0.00  & $\gamma_{AB}$ & 4.41 & eV ${\AA}^{2}$ \\
			$\gamma_{AC}$ &  0.00  &$\gamma_{AD}$&1.01  & eV ${\AA}^{2}$ \\
			$\chi_{AB} $  &  2.41  & 	$\chi_{AC}$ & 2.84& eV ${\AA}$ \\
		\end{tabular}
		\caption{\sf  Structure factor coefficients.}\label{bc}
	\end{table}\\
	In order to study the scattering of charge carriers in phosphorene, we consider 
	the following double barrier 
	\begin{equation}
		\label{eq 2}
		V_j(x)=\left\{\begin{array}{llll}
			{V_{1,5}=0} \ \ & \mbox{if} & {|x| > d_2}, \\
			{V_{2}} \ \ & \mbox{if} & {-d_2< x < -d_1}, \\
			{V_{3}} \ \ & \mbox{if} & {|x| < d_1}, \\
			{V_{4}}  \ \ & \mbox{if} & {d_1 < x < d_2}. \\
		\end{array}\right.
	\end{equation}
where $ V_{1,2,3} $ and $ d_{1,2,3} $ are positive potential parameters  characterizing the
double barrier structure shown in	
	Figure \ref{pprofil} 
	with
	$j$  labeling the five regions.
	
	According to the possible profile,  the 
	Hamiltonian describing the four regions $1,2,4, 5$ can be written as
	\begin{equation}
		H_{j}(k)=H^{+}(k)+V_{j}(x)\mathbb{I}_2
	\end{equation}
	and as for
	region $3$ we have 
	\begin{equation}\label{region3}
		H_{3}(k)=H^{+}(k)+V_{3}(x)\mathbb{I}_2+\Delta \sigma_z
	\end{equation}
	where $\sigma_z$ is the Pauli
	matrix and  $\Delta$
	is the involved  gap.
	Then,
	the dispersion relations  can be calculated to end up with the energies in regions $j=(1,2,4,5)$
	\begin{align}
		E_{j}=V_j+u_{0}+\eta_{x}k_{jx}^{2}+\eta_{y}k_{y}^{2}+s_{j}\sqrt{\left(\delta+ \gamma_{x} k_{jx}^{2}+\gamma_{y} k_{y}^{2}\right)^2+\left(\chi k_{jx}\right)^2}\label{wavevector10}
	\end{align}
as well as in region 3
	\begin{align}
	E_{3}=V_3+u_{0}+\eta_{x}k_{jx}^{2}+\eta_{y}k_{y}^{2}+s_{3}\sqrt{\left(\delta+ \gamma_{x} k_{jx}^{2}+\gamma_{y} k_{y}^{2}\right)^2+\left(\chi k_{jx}\right)^2+\Delta^2}\label{wavevector11}
\end{align}
	with  $s_j=\text{sign}\left(E-V_j-u_{0}-\eta_{x}k_{jx}^{2}-\eta_{y}k_{y}^{2}\right)$
and $s_3=\text{sign}\left(E-V_3-u_{0}-\eta_{x}k_{jx}^{2}-\eta_{y}k_{y}^{2}\right)$.
	Note that, 
	 the term proportional to $ k_{jx}^{2}$ in the energy spectrum can be ignored
	 because 
	 at low momenta we have  $u_{0}, \delta, k_{jx}\gg k_{jx}^{2}$ and  $ k_{jx}^{2}<1$ \cite{ref17}.
	The longitudinal wave vectors $k_{jx}$ corresponding to \eqref{wavevector10} and \eqref{wavevector11} are given by
	\begin{align}
&		k_{jx}=\chi^{-1}\sqrt{\left(E_{j}-V_j-u_{0}-\eta_{y}k_{y}^{2}\right)^2-\left(\delta+\gamma_{y} k_{y}^{2}\right)^2}\label{wavevector12} 
\\
&
		k_{3x}=\chi^{-1}\sqrt{\left(E_{3}-V_3-u_{0}-\eta_{y}k_{y}^{2}\right)^2-\left(\delta+\gamma_{y} k_{y}^{2}\right)^2-\left(\Delta\right)^2}\label{wavevector13}
	\end{align}
	
\begin{figure}[htbp]
		\centering{\includegraphics[width=2.9in]{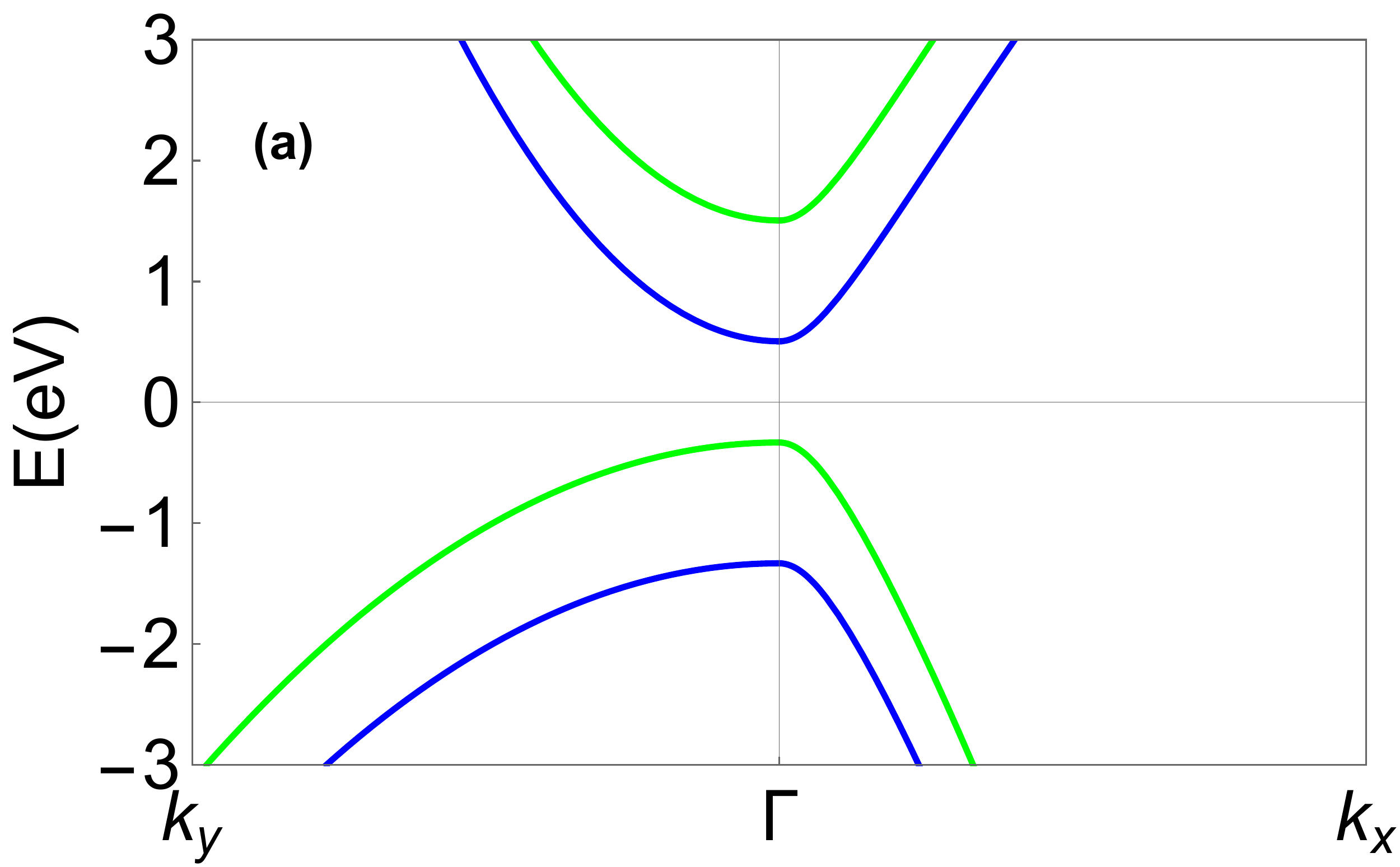}\ \ \ \includegraphics[width=2.9in]{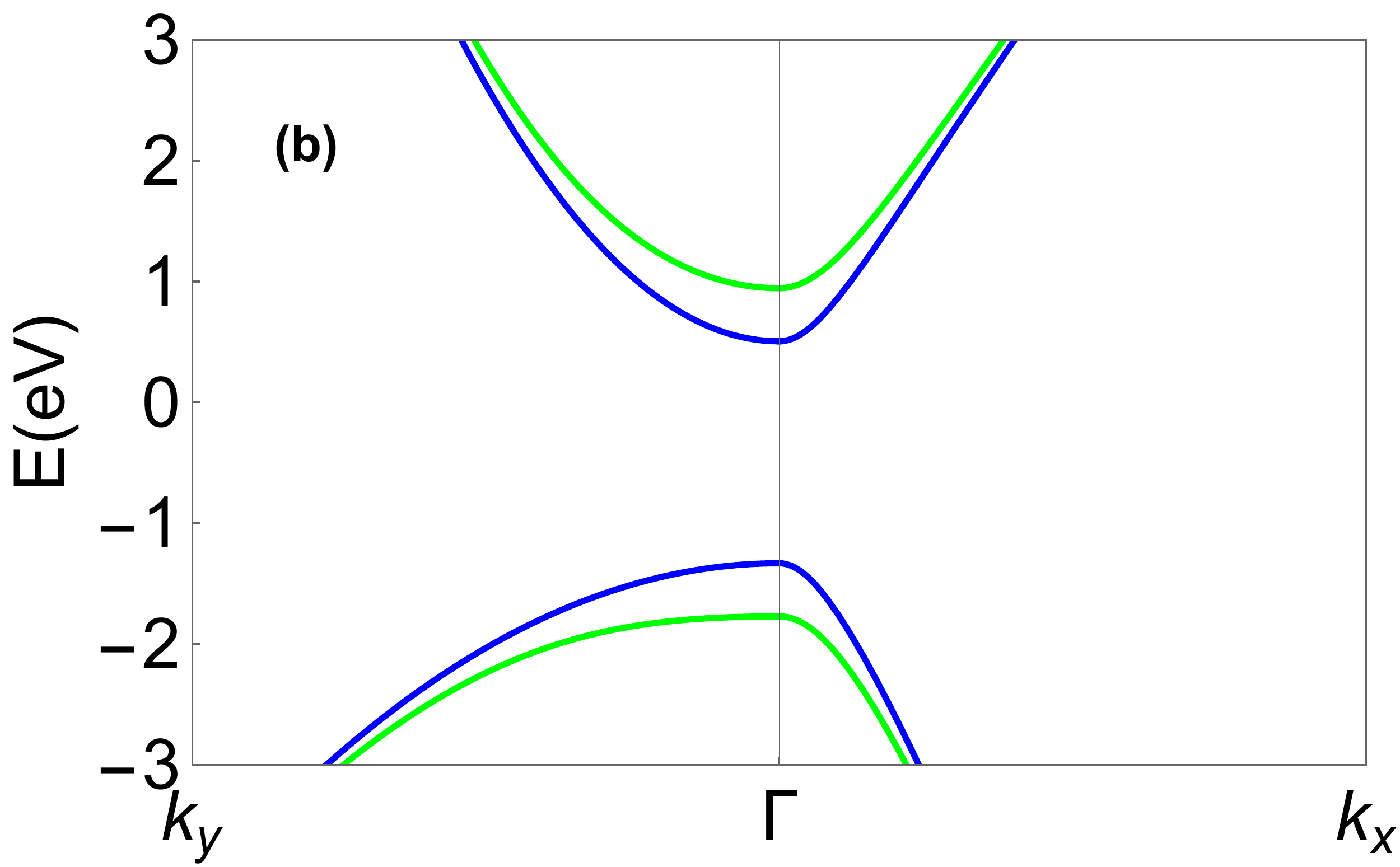}}
		\caption{{The energy eigenvalues outside and inside the
				barrier regions as a function of the wave vector $k_y$ and
				$k_x$. (a):   $V_3=0$ eV and $\Delta=0$ eV (green lines),   $V_3=1$ eV and $\Delta=0$ eV (blue lines). (b):  $V_3=0$ eV and $\Delta=1$ eV (green lines),  $V_3=0$ eV and $\Delta=0$ eV (blue lines).  }}\label{pote}
	\end{figure}

	We note that the energy spectrum is linear (Dirac like) in $k_x$ but
	parabolic (Schr\"odinger like) in $k_y$. Consequently, the nature
	of the quasiparticle transport and the effect of a potential barrier depends crucially on the its orientation in the $xy$-plane. 
	In addition, by increasing the potential height $V_3$, the energy bands are shifted upwards also by $V_3$, as shown in Figure \ref{pote}(a).
	Moreover, we can clearly see the effect of the energy gap, as shown in Figure \ref{pote}(b), that increase the gap between the valence and conduction bands.

	The associated  eigenspinors to \eqref{wavevector10} and \eqref{wavevector11} can be determined by solving the eigenvalue equation $H_j\Phi_j=E_j\Phi_j$. This process yields
	the solutions in five regions 	
	\begin{align}
		\label{eq3}
		&\Phi_{\sf 1}=  \left(
		\begin{array}{c}
			{1} \\
			{z_{1}} \\
		\end{array}
		\right) e^{i(k_{1x}x)} + r\left(
		\begin{array}{c}
			{1} \\
			{z_{1}^{-1}} \\
		\end{array}
		\right) e^{i(-k_{1x}x)}
		\\
		&\label{eq4}
		\Phi_{\sf 2}= a \left(
		\begin{array}{c}
			{1} \\
			{z_{2}} \\
		\end{array}
		\right) e^{i(k_{2x}x)} +b \left(
		\begin{array}{c}
			{1} \\
			{z_{2}^{-1}} \\
		\end{array}
		\right) e^{i(-k_{2x}x)}
		\\
		& \label{eq 7}
		\Phi_{\sf 3}= c \left(
		\begin{array}{c}
			{\sigma} \\
			{\tau z_{3}} \\
		\end{array}
		\right) e^{i(k_{3x}x)} +d \left(
		\begin{array}{c}
			{\sigma} \\
			{\tau z_{3}^{-1}} \\
		\end{array}
		\right) e^{i(-k_{3x}x)}
		\\
		&\label{eq5}
		\Phi_{\sf 4}= e \left(
		\begin{array}{c}
			{1} \\
			{z_{4}} \\
		\end{array}
		\right) e^{i(k_{4x}x)} +f \left(
		\begin{array}{c}
			{1} \\
			{z_{4}^{-1}} \\
		\end{array}
		\right) e^{i(-k_{4x}x)}
		\\
		&\label{eq6}
		\Phi_{\sf 5}= t \left(
		\begin{array}{c}
			{1} \\
			{z_{5}} \\
		\end{array}
		\right) e^{i(k_{5x}x)}
	\end{align}
	where we have set 
	$z_{j}=s_{j}e^{-i\theta_{j}}$ with 
	$s_{j}={\mbox{sign}}{\left(E-V_j-u_{0}-\eta_{y}k_{y}^{2}\right)}$ and the involved parameters are given by 	
	\begin{align}
		&\theta_{j}=\arctan\left( \frac{\chi k_{jx}}{\delta+\gamma_{y} k_{y}^{2}}\right)\\
	&	\sigma=\sqrt{1+\frac{s_3 \Delta}{\sqrt{\Delta^2+(k_{3x})^2}}}\\
	&
		\tau=\sqrt{1-\frac{s_3 \Delta}{\sqrt{\Delta^2+(k_{3x})^2}}}
	\end{align}

\section{Transmission and conductance}
	
	Next we will calculate the transmission probability of electrons across the  potential barriers in our phosphorene system. In doing so, it is convenient to use the matrix formalism. Using the boundary conditions at interfaces $x= -d_2$, $-$$d_1$, $d_1$, $d_2$ demanding the eigenspenors continuities
	\begin{align}
	\Phi_{\sf 1} (-d_2)=\Phi_{\sf 2} (-d_2), \qquad 
	\Phi_{\sf 2} (-d_1)=\Phi_{\sf 3} (-d_1), \qquad
	 \Phi_{\sf 3} (d_1)=\Phi_{\sf 4} (d_1), \qquad 
	 \Phi_{\sf 4} (d_2)=\Phi_{\sf 5} (d_2) 
	\end{align}
	which can be mapped as
	\begin{align}
	&	M_{1}[-d_2]\left(
		\begin{array}{c}
			{1} \\
			{r} \\
		\end{array}
		\right)=M_{2}[-d_2]\left(
		\begin{array}{c}
			{a} \\
			{b} \\
		\end{array}
		\right)\\
		&
	M_{2}[-d_1]\left(
	\begin{array}{c}
		{a} \\
		{b} \\
	\end{array}
	\right)	
	=M_{3}[-d_1]\left(
	\begin{array}{c}
		{c} \\
		{d} \\
	\end{array}
	\right)\\
	&
	M_{3}[d_1]\left(
	\begin{array}{c}
		{c} \\
		{d} \\
	\end{array}
	\right)
	=M_{4}[d_1]\left(
	\begin{array}{c}
		{e} \\
		{f} \\
	\end{array}
	\right)\\
	&
	M_{4}[d_2]\left(
	\begin{array}{c}
		{e} \\
		{f} \\
	\end{array}
	\right)=M_{5}[d_2]\left(
	\begin{array}{c}
		{t} \\
		{0} \\
	\end{array}
	\right)
	\end{align}
and then after a simple manipulation we get
	\begin{equation}\label{eq47}
		\left(
		\begin{array}{c}
			{1} \\
			{r} \\
		\end{array}
		\right)=M\left(
		\begin{array}{c}
			{t} \\
			{0} \\
		\end{array}
		\right)
	\end{equation}
where  the introduced transfer matrix is given by
	\begin{align}
		\label{matrix}
		M &= M_{1}^{-1}[-d_2]\ M_{2}[-d_2]\
		M_{2}^{-1}[-d_1] \ M_{3}[-d_1]\ M_{3}^{-1}[d_1]\ M_{4}[d_1]\ M_{4}^{-1}[d_2]\ M_{5}[d_2]\\
		&=
		\begin{pmatrix}
		M_{11}& 	M_{12}\\
		M_{21}& M_{22}
		\end{pmatrix}
	\end{align}

Since the wave vector incident and transmitted waves is the same, 
therefore, the transmission probability can be obtained as
\begin{equation}
	T = |t|^2= \frac{1}{|M_{11}|^2}
\end{equation}
where $t$ is given by
\begin{align}
	t=\frac{16\tau\sigma s_1 s_2 s_3 s_4 \sin\theta_1\sin\theta_2\sin\theta_3\sin\theta_4 e^{i\left(d_2(-2k_{1x}+k_{2x}+k_{4x})+d_1(k_{2x}+2k_{3x}+k_{4x})\right)}}
	{e^{4id_1k_{3x}} A_1 A_2
	+B_1 B_2} 
\end{align}
and the parameters are 
\begin{align}
	& A_1=(z_{1}-z_{4}^{-1})(z_{4}\sigma-z_{3}\tau) e^{2id_1k_{4x}}+ (z_{1}-z_{4})(z_{4}^{-1}\sigma-z_{3}\tau) e^{2id_2k_{4x}}\\ 
	& A_2=(z_{1}^{-1}-z_{2}^{-1})(z_{2}\sigma-z_{3}^{-1}\tau) e^{2id_2k_{2x}}+
	(z_{1}^{-1}-z_{2})(z_{2}^{-1}\sigma-z_{3}^{-1}\tau) e^{2id_1k_{2x}}\\
	& B_1=(z_{1}-z_{4})(z_{4}^{-1}\sigma-z_{3}^{-1}\tau) e^{2id_2k_{4x}}+  (z_{1}-z_{4}^{-1})(z_{4}\sigma-z_{3}\tau) e^{2id_1k_{4x}}\\
	& B_2=(z_{1}^{-1}-z_{2}^{-1})(z_{2}\sigma-z_{3}\tau) e^{2id_2k_{2x}}+ (z_{1}^{-1}-z_{2})(z_{2}^{-1}\sigma-z_{3}\tau) e^{2id_1k_{2x}}.
\end{align}

	
To complete our study, we investigate an important physical quantity relevant to double barrier structure in phosphorene that is the conductance. Then,
		using the Landauer-Buttiker formula \cite{ref17} and obtained transmission probability we get
	\begin{equation}
		G=G_0 \int_{-k_y^{\text{max}}}^{k_y^{\text{max}}} \frac{d k_y}{2\pi}T
	\end{equation}
	where $G_0 = \frac{e^2 L_y}{\hbar}$ is the unit conductance, $L_y$ is the width of the sample in the $y$-direction and $k_y^{\text{max}}$ denotes the maximum transverse momenta. 
	In next section, we will study numerically these two quantities. This will help us understand the effect of various physical parameters on the transmission and the conductance related to our double barrier structure.
	
	\section{Results and discussions}\label{Sec3}

\begin{figure}[H]
		\centering
		\includegraphics[width=8cm, height=5cm]{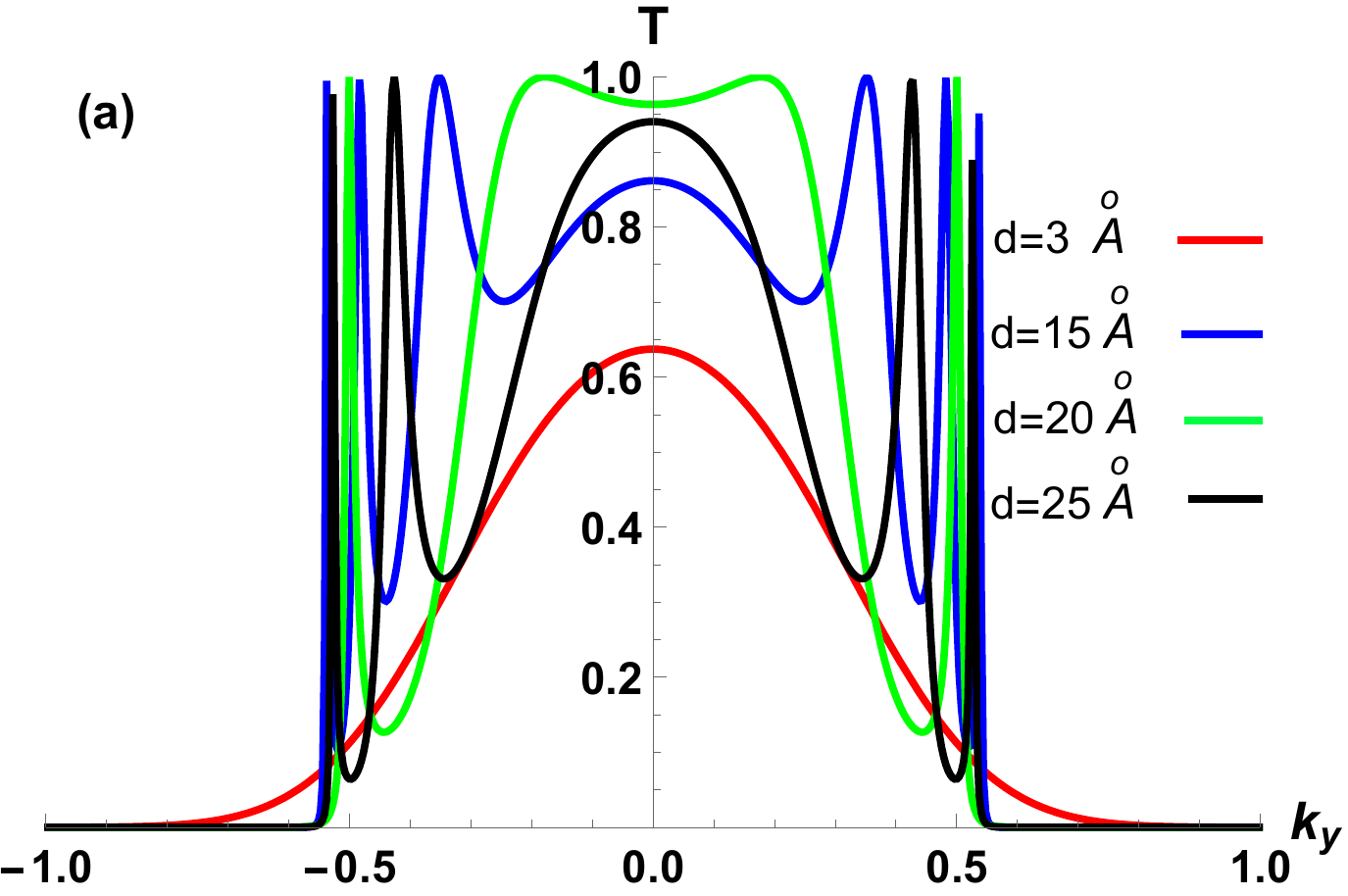}\ \ \ \ \
		\includegraphics[width=8cm, height=5cm]{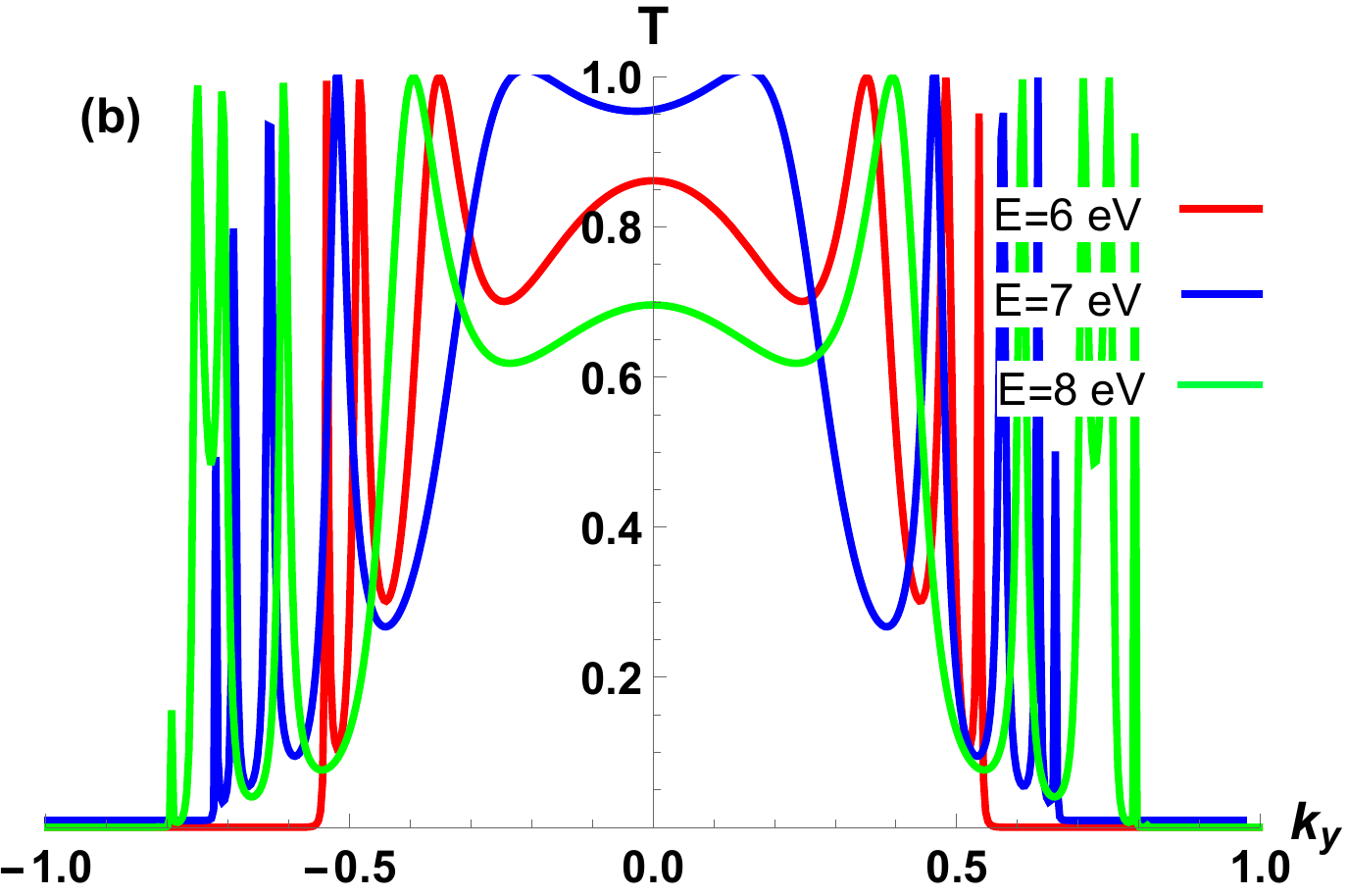}\\
		\caption{\sf  Transmission probability $T$ as a function of $k_y$, with $V_2= V_4 = 12$  eV , $V_3 = 4$  eV , $\Delta = 0$  eV and $d_2 = 2d_1 = d$ for different values of $d$ and $E$. (a) : $d = 3 \text{ \AA}$ (red line), $d = 15 \text{ \AA}$ (blue line), $d = 20\text{ \AA}$ (green line), $d = 25 \text{ \AA}$ (black line) and $E = 6$  eV. (b) : $E = 6$  eV (red line), $E = 7$  eV (blue line), $E = 8$  eV (green line) and $d = 25\text{ \AA}$.	}\label{Tky1}
	\end{figure}

	Figure \ref{Tky1} shows the transmission probability 
	as a function of the wave vector $k_y$ for different values of barrier width $d$ and incident energy $E$. The barrier heights are fixed at $V_2 = V_4 = 12$  eV, $V_3 = 4$~eV, and the gap $\Delta = 0$  eV.  $T$ is plotted for $d = 3 {\text{\AA}} $ (red line), 15$ {\text{\AA}} $ (blue line), 20 $ {\text{\AA}} $ (green line), 25 $ {\text{\AA}} $ (black line) in Figure \ref{Tky1}(a) 
	for $E = 6$ (red line), 7 (blue line) and $8$  eV
	(green line) in Figure \ref{Tky1}(b). 
	We observe that the transmission is bilaterally symmetrical with respect to the normal incidence, i.e. $k_y = 0$,
	and 
	exhibits a behavior similar to that observed for Dirac particle in graphene \cite{ref18,ref19,ref20}. It is important to note that in the case of graphene in the absence of a magnetic field, the perfect transmission situation appears very clearly in the case $k_y = 0$
	(signature of Klein tunneling), which is not the case in our system made of phosphorene. As shown in Figure \ref{Tky1}(a), 
	 the transmission through thin barriers is negligible compared to that for large $d$, such 
	 behavior is similar to that seen for the transmission thought a single barrier in monolayer phosphorene \cite{ref17}. From Figure \ref{Tky1}(b), we notice that
	the transmission vanishes for specific values that increase by increasing  $E$. Note that when $k_y$ exceed a critical value the transmission probability go to zero	(the transmission is blocked by the barrier) due to the evanescent nature of the states in the barrier.
	There exist several line-shaped peaks in the forbidden transmission region that are consequences of resonant transmission through the double barrier. The number of resonant peaks 
	changes with the change  of the energy value as can be seen in Figure \ref{Tky1}(b).
	
\begin{figure}[H]
		\centering
		\includegraphics[width=7cm, height=5cm]{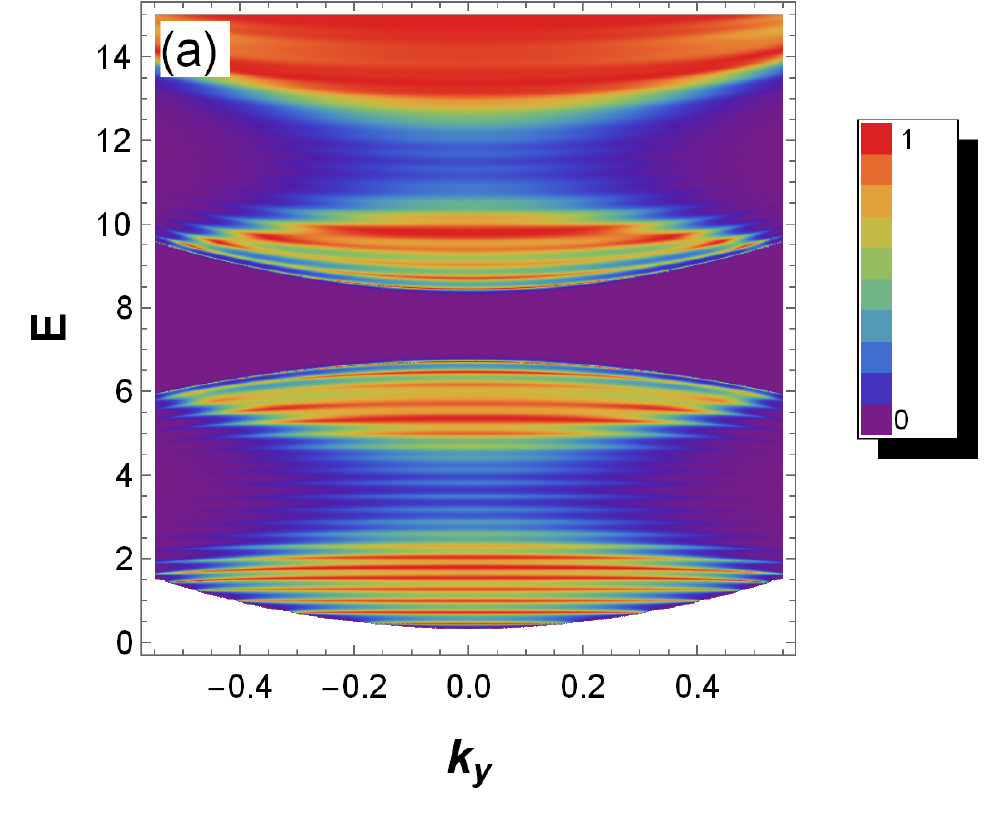}\ \ \ \ \
		\includegraphics[width=7cm, height=5cm]{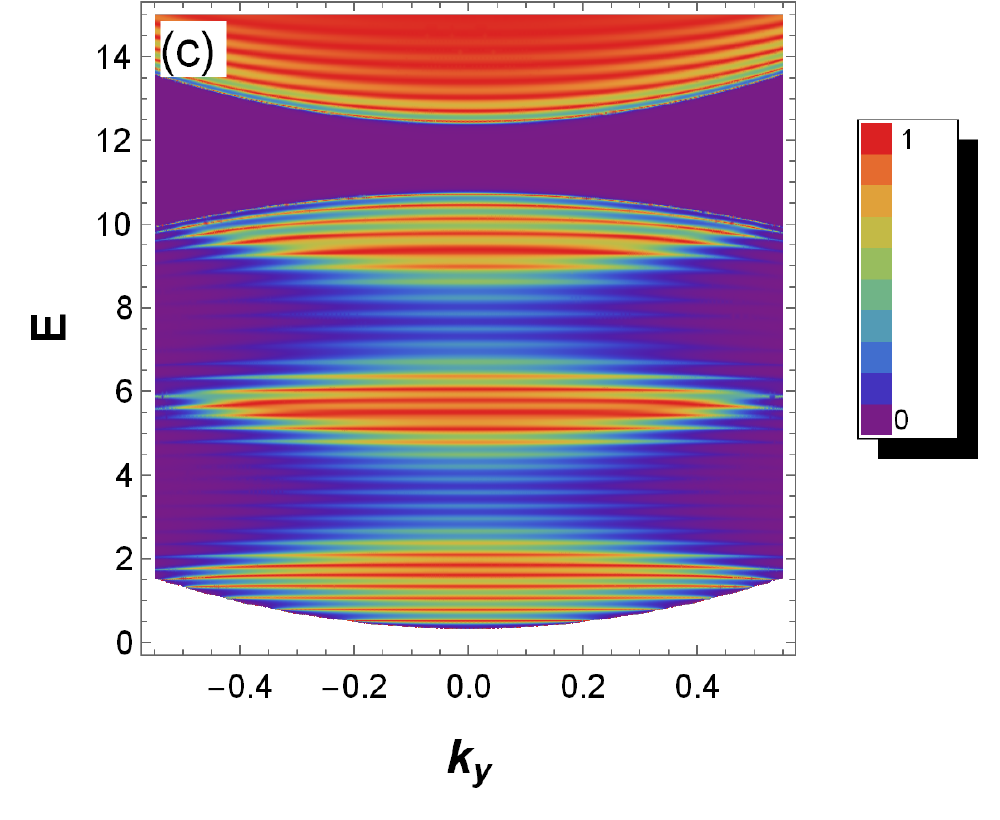}\\
		\includegraphics[width=7cm, height=5cm]{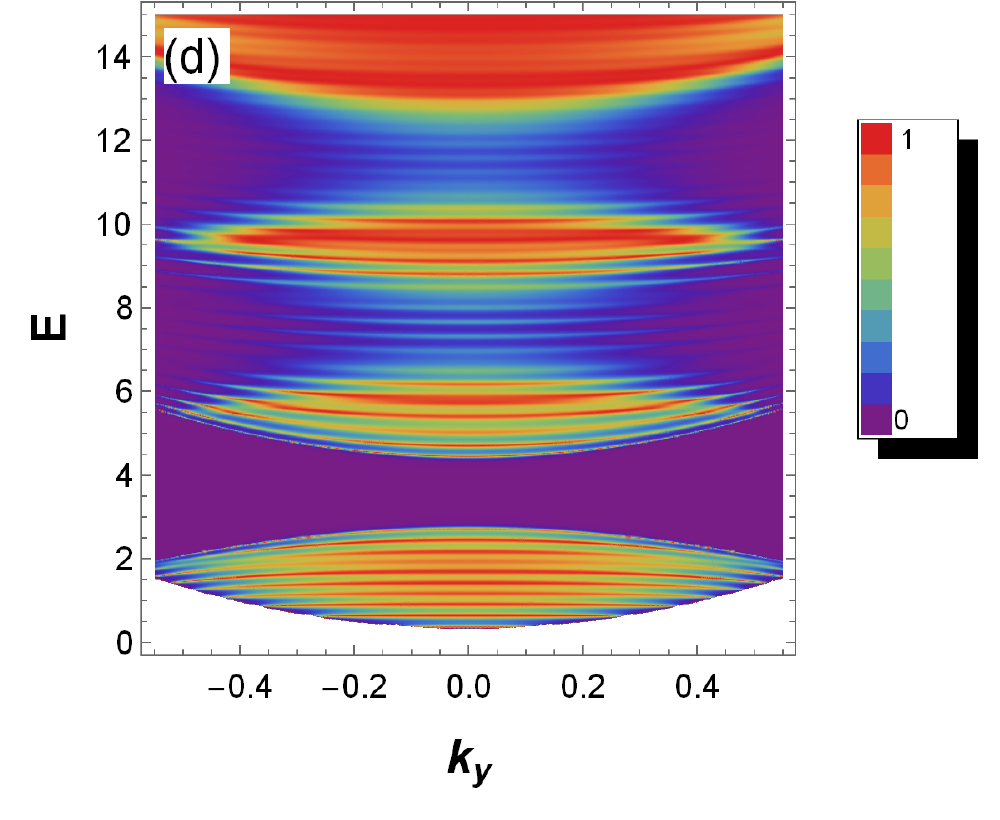}\ \ \ \ \
		\includegraphics[width=7cm, height=5cm]{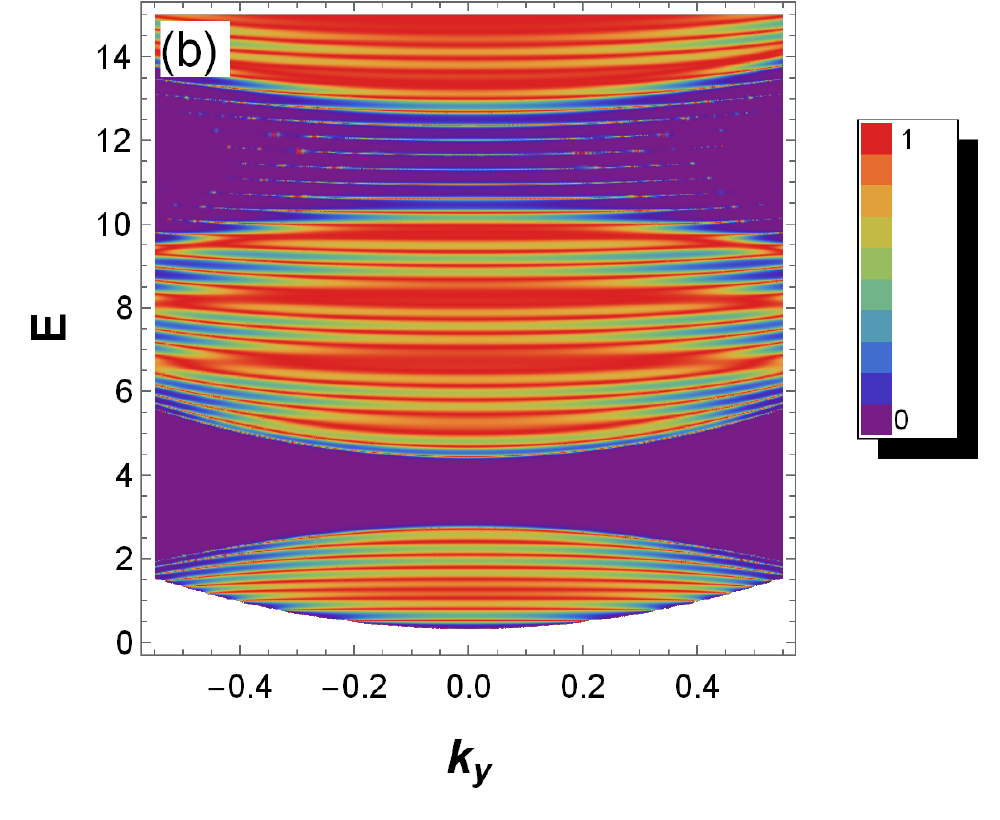}\\
		\caption{\sf Density plot of transmission probability $T$ as a function of the transverse wave vector $k_y$ and energy $E$, with $d_1 = 20\ \text{\AA}$, $d_2 = 30.3\ \text{\AA}$ and $\Delta = 0$  eV, for different barrier structures. (a): $V_2 = 4$ eV, $V_3 = 8$  eV and $V_4 = 12$  eV. (b): $V_2 = V_4 = 12$  eV and $V_3 = 4$  eV. (c): $V_2 = 4$  eV, $V_3 = 12$ eV and $V_4 = 8$  eV. (d): $V_2 = 8$ eV, $V_3 = 4$  eV and $V_4 = 12$  eV.
		}\label{Dtky}
	\end{figure}

In Figure \ref{Dtky}, we present 
the transmission probability as a function of the transverse wave vector $k_y$ and the energy $E$ for four configurations 
of the double barrier. It is clearly seen that there are different energy zones  characterizing the transmission. Indeed,  Figure \ref{Dtky}(a) corresponds to the case
$V_2 < V_3 < V_4$ and shows 
seven energy zones. 
At normal incidence, we observe that  the first, third and fifth zones are bounded by the energy intervals 
$0 < E < V_2 + u_0 - \delta$, $V_2 + u_0 + \delta < E < V_3 + u_0 - \delta$ and $V_3 + u_0 + \delta < E < V_4 + u_0 - \delta$, respectively, which contains oscillations (resonances) in the transmission. While the transmission is zero in the second, fourth and sixth zones $V_i +u_0-\delta < E < V_i +u_0 +\delta $ with $i = 2, 3$ and 4, respectively. It is important to note that the transmission displays sharp peaks inside the transmission gap around the point $E = V_2 + u_0$ and $E = V_4 + u_0$, but they are absent around the energy point $E = V_3 + u_0$. Finally, the seventh zone $E > V_4 + u_0 + \delta$ contains the usual high energy barrier oscillations and asymptotically goes to unity at high energy. It is important to note that, the transmission exhibit the same behavior for both cases  $V_2 < V_4 < V_3$ and $V_3 < V_2 < V_4$ as shown in Figure \ref{Dtky}(b,c).
Form Figure \ref{Dtky}(d) for  $V_2 = V_4 \ne V_3$, we observe that there are only five energy zones 
where in the first ($0 < E < V_2 +u_0-\delta$)
and third ($V_2 +u_0-\delta < E < V_4 +u_0 +\delta$) zones, the transmission exhibits an oscillatory behavior as a function of the incident energy. Moreover, the transmission displays sharp peaks inside the transmission gap around the point $E = V_2 +u_0 = V_4 +u_0$, that are absent around $E = V_3 +u_0$. These peaks can be attributed to the quasi-bound states formed in the double barrier structure
in  similar way to that obtained for graphene \cite{ref21}.

\begin{figure}[H]
		\centering
		\includegraphics[width=7cm, height=5cm]{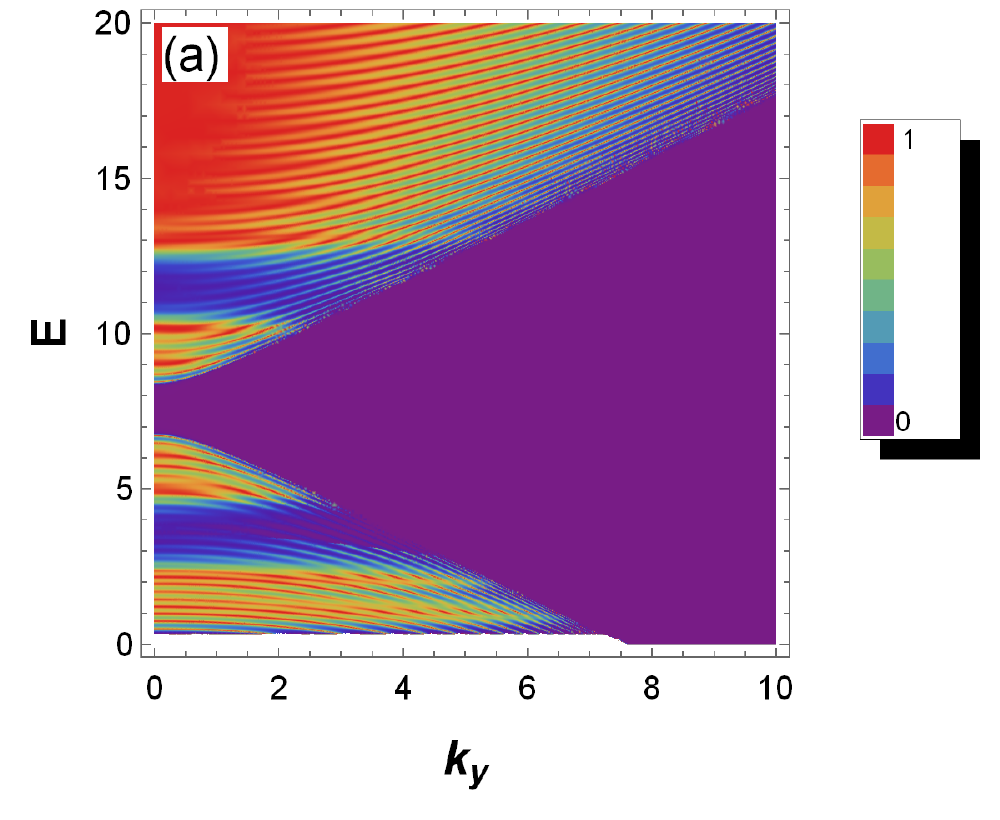}\ \ \ \ \
		\includegraphics[width=7cm, height=5cm]{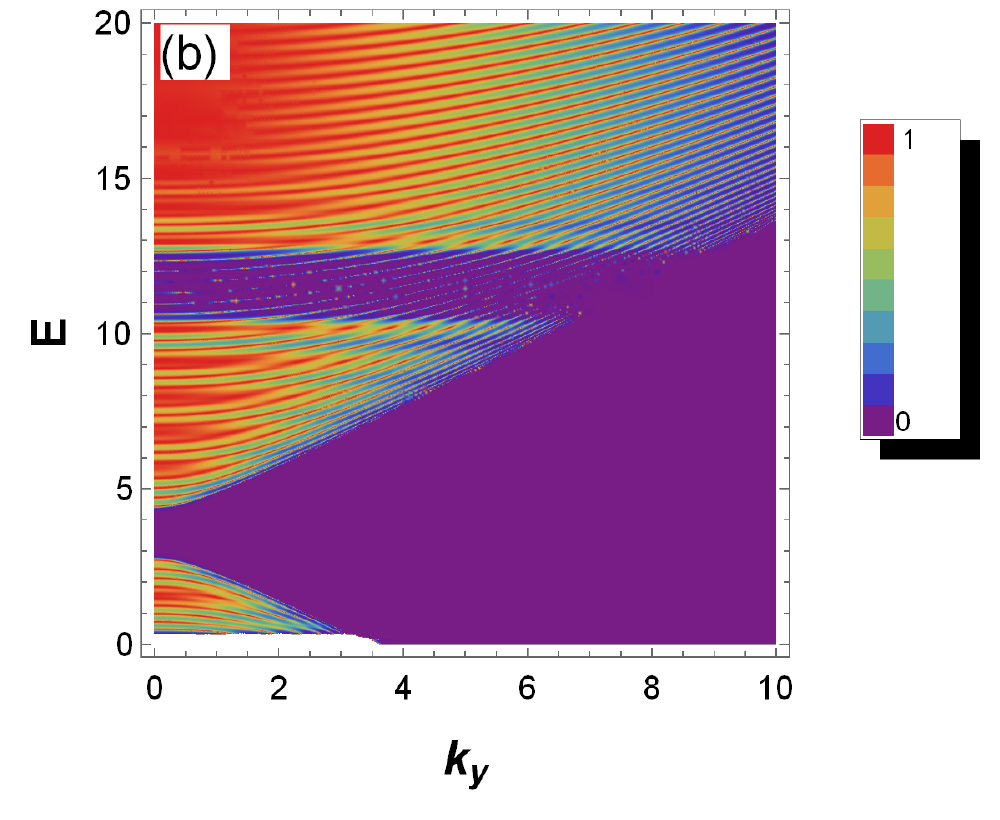}\\
		\includegraphics[width=7cm, height=5cm]{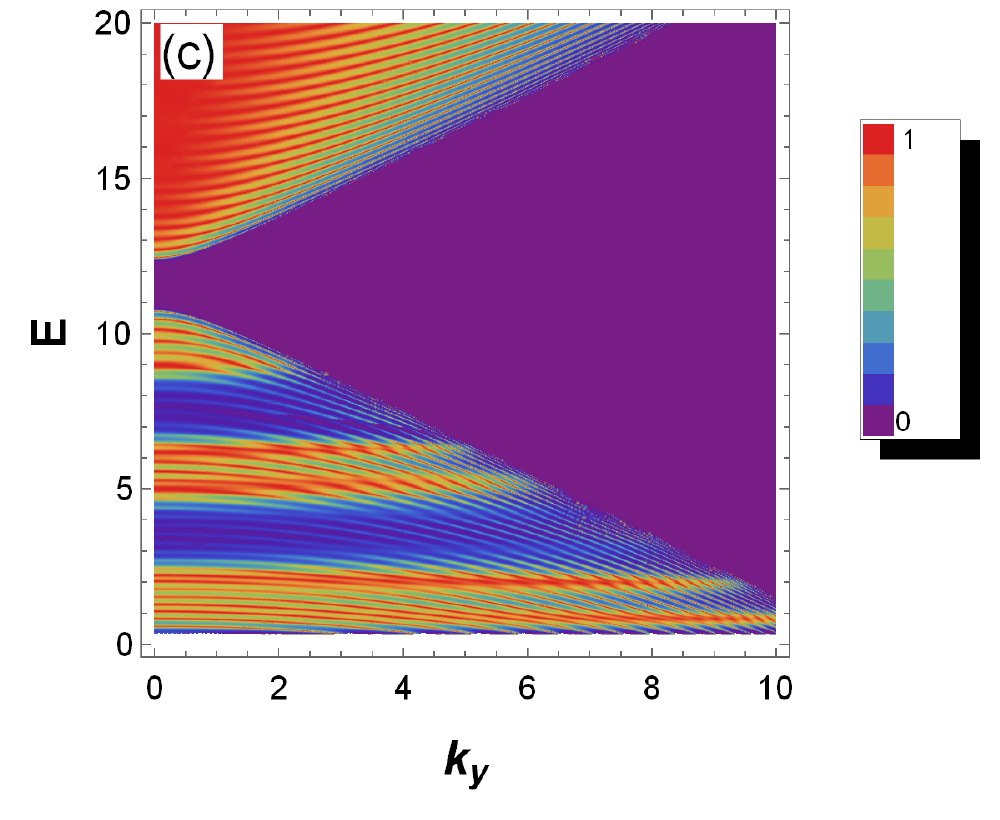}\ \ \ \ \
		\includegraphics[width=7cm, height=5cm]{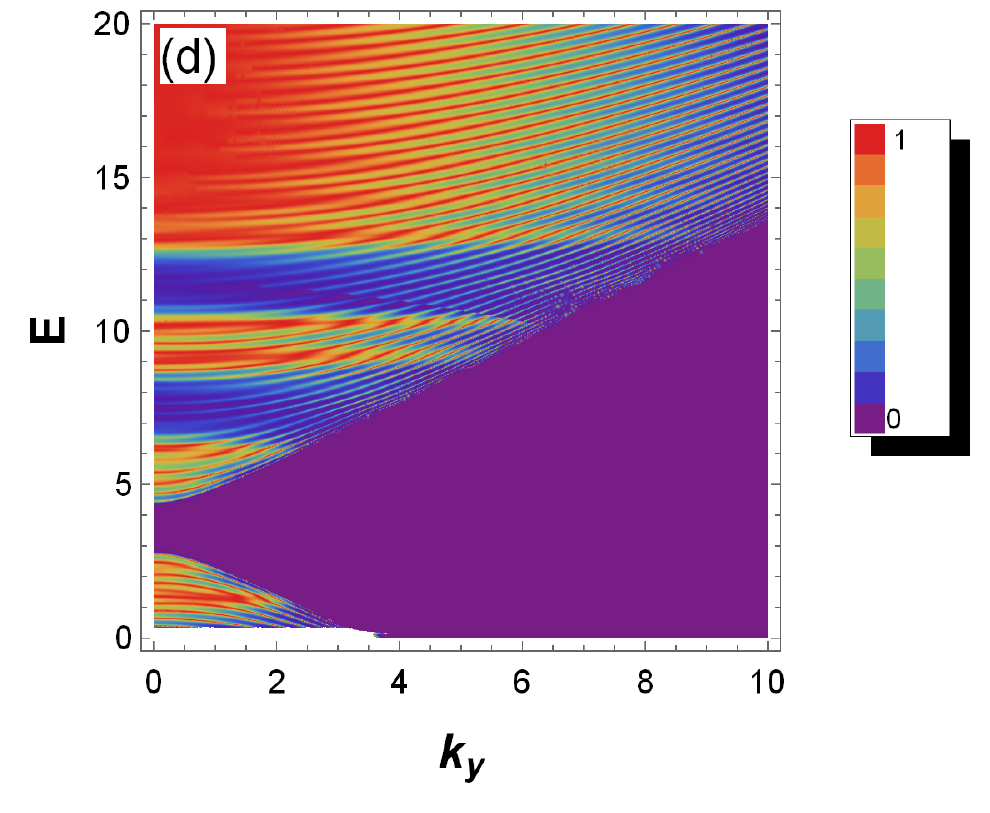}\\
		\caption{\sf Density plot of transmission probability $T$ as a function of energy gap $\Delta$ and incident energy $E$ with $d_1 = 20 \ \text{\AA}$, $d_2 = 30.3 \ \text{\AA}$ and $k_y = 0 \ \text{\AA}^{-1}$ for different barrier structures. (a): $V_2 = 4$  eV, $V_3 = 8$  eV and $V_4 = 12$  eV. (b): $V_2 = V_4 = 12$  eV and $V_3 = 4$ eV. (c): $V_2 = 4$  eV, $V_3 = 12$  eV and $V_4 = 8$  eV. (d): $V_2 = 8$  eV, $V_3 = 4$  eV and $V_4 = 12$  eV.
		}\label{DtkyDelta}
	\end{figure}

	We now investigate the effect of the introduced gap $\Delta$
	on the transmission probability $T$ in 
	Figure~\ref{DtkyDelta}, which
	has been performed by fixing the parameters $d_1 = 20 \ \text{\AA}$, $d_2 = 30.3 \ \text{\AA}$ and $k_y = 0$ for different barrier structures as in Figure \ref{Dtky}. We observe  that for  $\Delta = 0$, $T$ exhibits sharp peaks around $E = V_2 + u_0$ and $E = V_4 + u_0$, but they are absent around $E = V_3 + u_0$, which confirm what we found   in Figure~\ref{Dtky}. It is clearly seen that 
	as long as   $\Delta$ the transmission gap around $E = V_3 + u_0$ increases in all cases and  for large $\Delta$, the transmission oscillation starts to decrease until vanishing.

\begin{figure}[H]
		\centering
		\includegraphics[width=7cm, height=5cm]{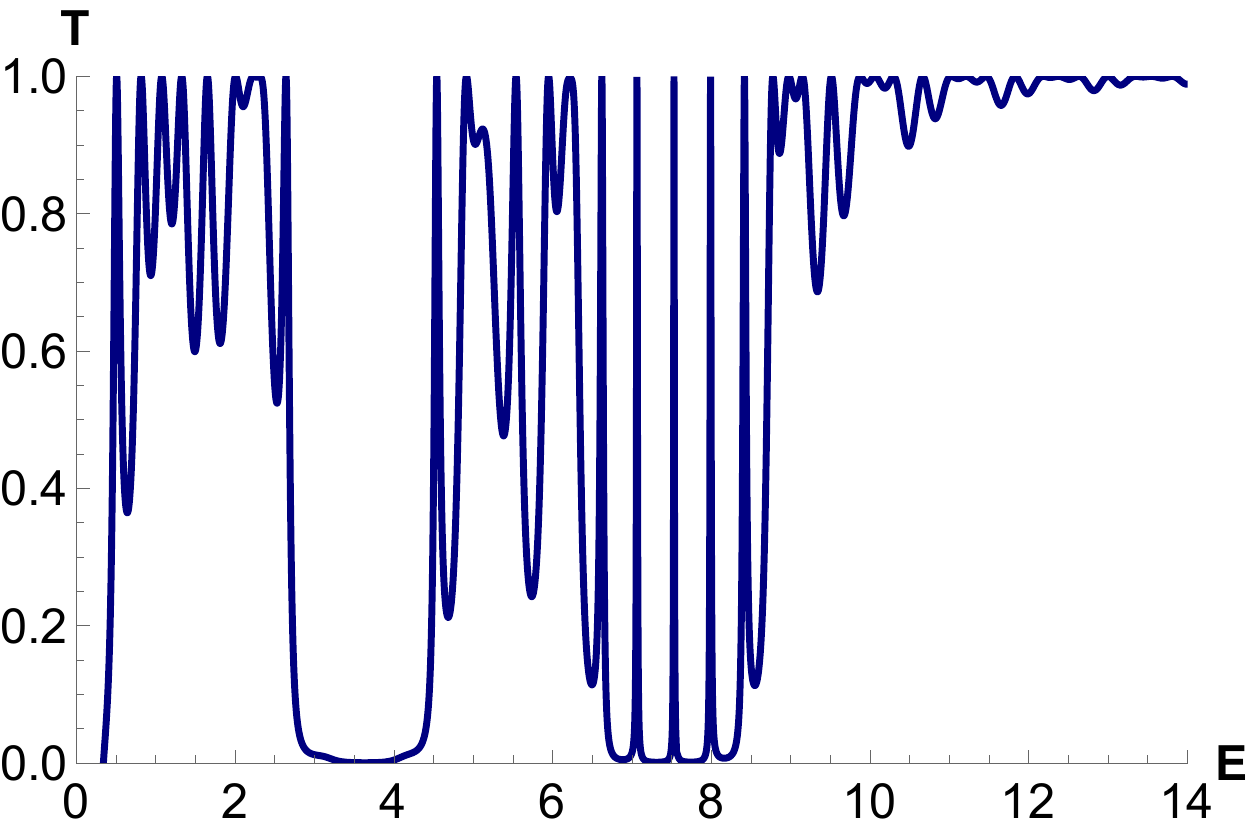}\ \ \ \ \
		\includegraphics[width=7cm, height=5cm]{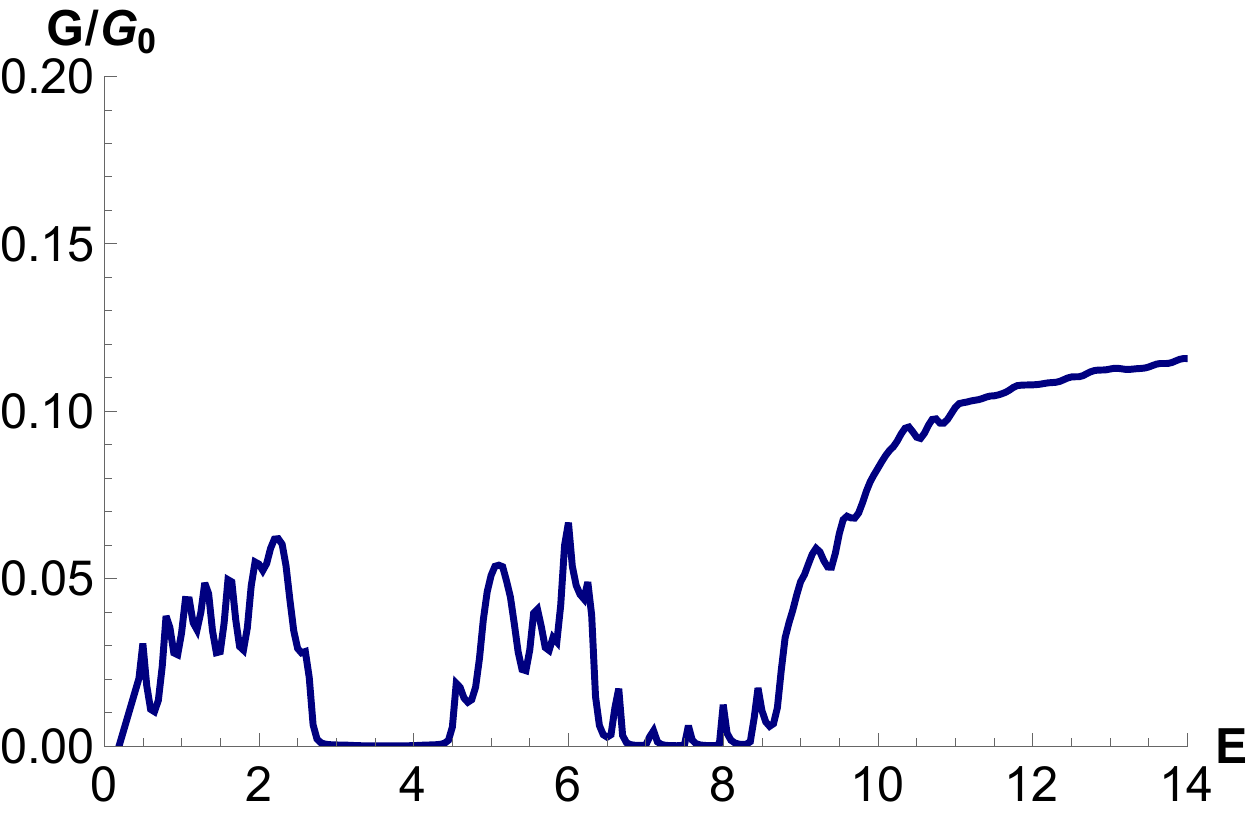}\\
		\caption{\sf Transmission probability and conductance as a function of the incident energy $E$ for $d_2 = 2 d_1  = d = 12 \ \text{\AA}$, $\Delta = 0$, $V_2 = V_4 = 8$ eV, $V_3 = 4$  eV and  $k_y = 0$. 
		}\label{TGE}
	\end{figure}

	It is worth while to investigate the conductance as a function of the incident energy together with transmission in Figure \ref{TGE}. 
	To analyze the conductance behavior we distinguish five zones such that in the first zone 
	($E<V_3 + u_0 -\delta$), the resonances in the transmission 
	correspond to peaks in the conductance. The second zone ($V_3 + u_0 - \delta < E < V_3 + u_0 +\delta $) shows a window where $T$ and $G/G_0$ are zero. In the third zone ($V_3 + u_0 + \delta < E < V_4 + u_0 - \delta$), the peaks in 
	conductance
	 have shoulders due to the presence of resonances in the transmission. We notice that the number of peaks is the same for the transmission and the conductance. The fourth ($V_4 + u_0 -\delta < E < V_4 + u_0 + \delta$) presents a bowl where $T$ and $G/G_0$ are mostly zero but contains resonances peaks. The fifth zone ($V_1 < E_1$) contains oscillations, the transmission converges to unity at high energies and conductance researches a maximum value.

\begin{figure}[H]
		\centering
		\includegraphics[width=7cm, height=5cm]{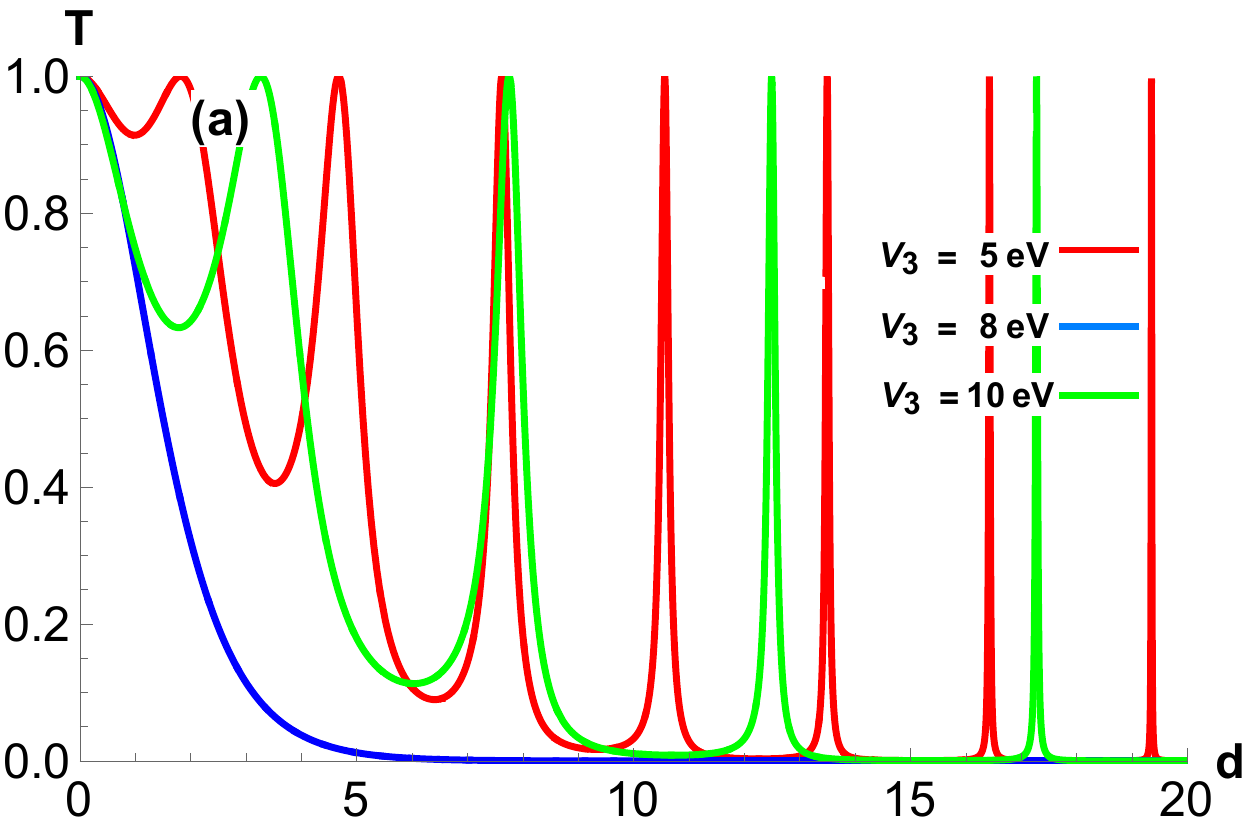}\ \ \ \ \ \ \ \
		\includegraphics[width=7cm, height=5cm]{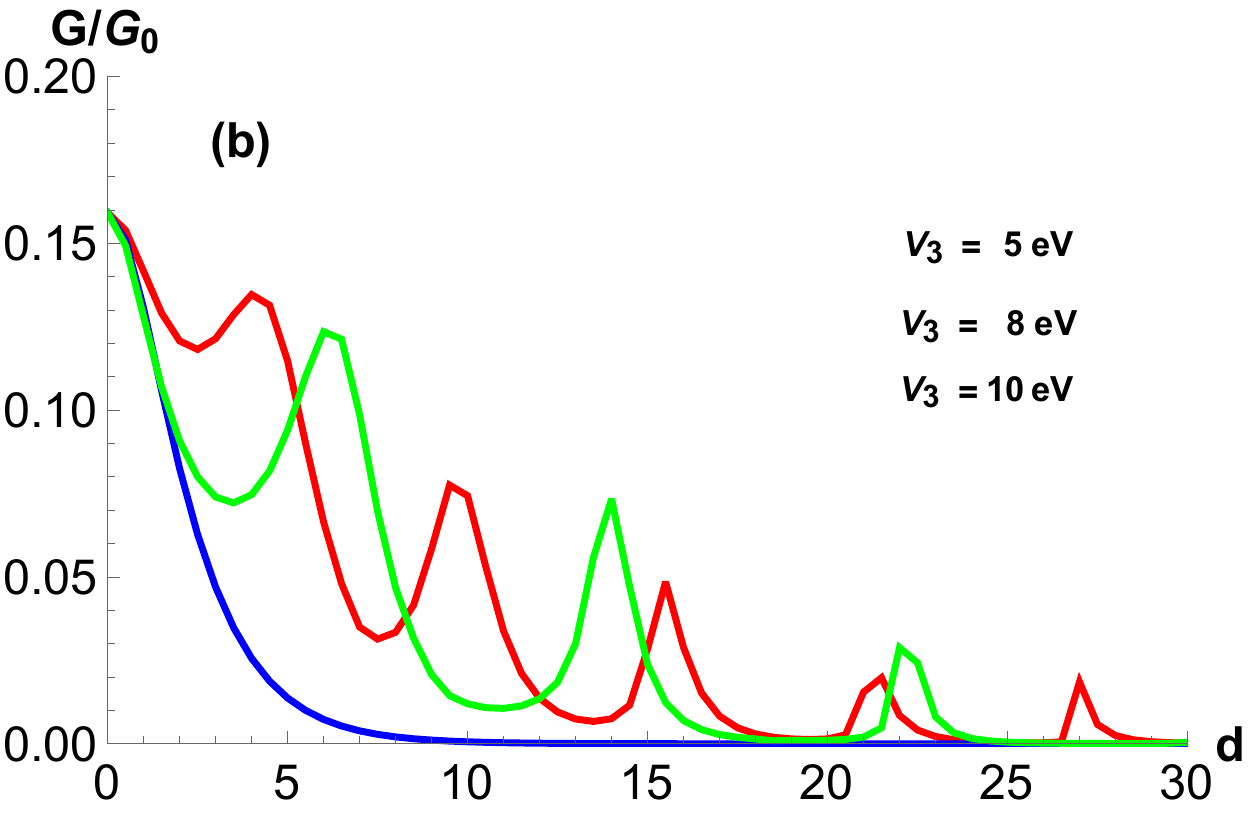}\\
		\caption{\sf  The transmission probability $T$ and conductance $G/G_0$ as a function of the barrier width $d$ with $d_1=d/2$, $d_2=d$ for $\Delta=0$ eV, $E=7.5$  eV and $k_y=0$. $V_2=V_4=8$  eV, $V_3=5$  eV (red solid line), $V=8$  eV (blue solid line),  $V=10 eV$ (green solid line). 
		}\label{fig8}
	\end{figure}

	Figure \eqref{fig8} is intended to see the influence of the barrier  width $d$ for different values of the central potential $V_3$. It is clearly seen  that the transmission $T$ and  conductance $G/G_0$ are either a decaying or an oscillatory  
	depending on the value of $V_3$. For the range $E < V_4 + u_0 < V_3 + u_0$, $T$ and $G/G_0$ display oscillatory behavior with the increase of $d$ in similar way  
	to that of massless Dirac like electrons as seen in graphene \cite{ref21}. However in contrast to graphene, $T$ and $G/G_0$ show a sharp decay as a function of $d$ when $E < V_3+u_0 < V_4+u_0$. For $V_3+u_0 < E < V_4+u_0$, $T$ and $G/G_0$ show an oscillatory behavior
	similar 
	to 
	single and multiple barrier structure in monolayer phosphorene \cite{ref17}.
	
	\section{Conclusion}

	We have investigated the transport properties of charge carriers transmitted through monolayer phosphorene with double barriers. We have started by formulating the Hamiltonian model describing our system  and getting the associated energy bands. We have shown that the anisotropic properties arise from the difference between $\Gamma-X$ and $\Gamma-Y$ dispersions
	corresponding to the armchair and zigzag directions, respectively. 
	Subsequently, by
	using the transfer matrix method, we have calculated the transmission 
	as well as  the conductance in terms of the physical parameters. 
	
	Our numerical analysis showed that
	the transmission displays sharp pics inside the transmission gap around $E = V_4 + u_0$ and $E = V_2 + u_0$, which are 
	 absent around $E = V_3 + u_0$. These peaks can be attributed to the quasi-bound states formed in the double barrier structure. This behavior was observed in monolayer graphene with double barrier structure.  Moreover, we have found that the transmission is bilaterally symmetrical with respect to the normal incidence $k_y$ and  
	 there is no signature of the Klein tunneling 
	 contrary to graphene. 
	It was argued  with the  increase of $k_y$ the transmission vanishes for some specific values depending on the incident energy. We have observed that the transmission through thin barriers is negligible compared to that  in single barrier for large  width $d$. It was shown that when $k_y$ exceed a critical value the transmission is blocked by the barrier, which is due to the evanescent nature of the states in the barrier. We have seen that many line-shaped peaks appeared in the forbidden transmission region whose number changes with the change of energy that are the consequences of resonant transmission through the double barrier.

	Finally, we have shown that the transmission and the conductance display oscillatory behavior as a function of the barrier width $d$ for both cases $E < V_4 + u_0 < V_3 + u_0$ and $V_3 + u_0 < E < V_4 + u_0$, but  sharp decay for $E < V_3 + u_0 < V_4 + u_0$ similar to that obtained for single barrier in phosphorene. It was found that the conductance as a function of the incident energy presents a behavior similar to the transmission one. It was argued that for $E<V_3 + u_0 - \delta$ the resonances in  transmission correspond to peaks in
conductance. Moreover, around $E= V_3 + u_0 $ and $E= V_4 + u_0 $, we have found that the conductance is zero, while the peaks in the conductance around $E= V_4 + u_0 $ are  due to the presence of resonances in the transmission probability.


\end{document}